\documentclass{PoS}
\usepackage{url}

\title{Exploring walking behavior in SU(3) gauge theory with 4 and 8 HISQ quarks}

\ShortTitle{Exploring walking behavior in SU(3) gauge theory with 4 and 8 HISQ quarks }

\author{Yasumichi Aoki$^a$, \, Tatsumi Aoyama$^a$, \, Masafumi Kurachi$^a$, \, Toshihide Maskawa$^a$, \, \speaker{Kei-ichi Nagai} \thanks{E-mail: keiichi.nagai@kmi.nagoya-u.ac.jp}\, $^a$, \, Hiroshi Ohki$^a$, \, Akihiro Shibata$^b$, \, Koichi Yamawaki$^a$  and \, Takeshi Yamazaki$^a$ 

\hspace{55mm} LatKMI Collaboration\\
        $^a$Kobayashi-Maskawa Institute for the Origin of Particles and the Universe (KMI), 
        Nagoya University, Nagoya, 464-8602, Japan\\
        $^b$Computing Research Center, High Energy Accelerator Research Organization (KEK), 
        Tsukuba, 305-0801, Japan \\
}

\abstract{We present the report of the LatKMI collaboration on the lattice QCD simulation 
for the cases of 4 and 8 flavors. 
The Nf=8 in particular is interesting from the model-building point of view: 
The typical walking technicolor model with the large anomalous dimension 
is the so-called one-family model (Farhi-Susskind model). 
Thus we explore the walking behavior in LQCD with 8 HISQ quarks 
by comparing with the 4-flavor case (in which the chiral symmetry is spontaneously broken). 
We report preliminary results on the spectrum, 
analyzed through the chiral perturbation theory 
and the finite-size hyperscaling, 
and we discuss the availability of the Nf=8 QCD to the phenomenology.
}

\FullConference{The 30 International Symposium on Lattice Field Theory - Lattice 2012,\\
		June 24-29, 2012\\
		Cairns, Australia}

\begin{document}

\section{Introduction}
\label{sec:intro}

The origin of mass is the most urgent issue of the particle physics today.
One of the candidates for the theory 
beyond the Standard Model 
is the walking technicolor 
which is the strongly coupled gauge theory 
having a large anomalous dimension 
$\gamma_m \simeq 1$  
and approximate scale invariance 
due to the almost  non-running (walking) coupling \cite{Yamawaki:1985zg,Akiba:1985rr}.
The walking behavior is in fact  realized in the QCD 
with  large number of (massless) flavors  $N_f$ 
which possesses 
Caswell-Banks-Zaks infrared fixed point (IRFP)  \cite{beta}
in the two-loop beta function. 
The exact IRFP would be  washed out by  the dynamical generation of a quark mass $m$ 
in the very infrared region $\mu<m$
for $N_f < N_f^{cr}$, 
$N_f^{cr}$ being the critical number. 
However,  for  $N_f$ very close to $N_f^{cr}$,  
$m$ could be much smaller than the intrinsic scale  $\Lambda$  ($\gg  m $), 
an analogue of $\Lambda_{\rm QCD}$, 
beyond which the coupling runs as the asymptotically free theory,  
so that the coupling remains almost walking
for the wide infrared region $m<\mu<\Lambda$ 
as a remnant of the would-be IRFP. 
The case $N_f >N_f^{cr}$ is called conformal window, 
although conformality is broken 
in the ultraviolet  asymptotically free region beyond $\Lambda$.

Although the results from the two-loop 
and ladder approximation 
of Schwinger-Dyson equation analysis~\cite{Maskawa:1974vs}
are very suggestive,  
the relevant dynamics is obviously of  non-perturbative nature,
we would need fully non-perturbative studies. 
Among others the lattice simulations developed in the lattice QCD 
would be the most powerful tool for that purpose.
The above two-loop and ladder studies suggest 
that the walking theory if existed would be in between $N_f=8$ and $N_f=12$.
The $N_f=8$ in particular is interesting 
from the model-building point of view: 
The typical technicolor model \cite{Farhi:1980xs}  is the so-called 
one-family model (Farhi-Susskind model)  
which has a one-family of the colored techni-fermions (techni-quarks) 
and the uncolored one (techni-leptons)
corresponding to the each family of the SM quarks and leptons. 
Thus if the $N_f=8$ turns out to be a  walking theory, 
it would be a great message for the phenomenology 
to be tested by the on-going LHC. 

Since the pioneering works on the lattice~\cite{Iwasaki:1991mr,Appelquist:2007hu} 
were carried out,
a lot of groups have been doing lattice studies nowadays. 
(See Refs.~\cite{plenary} for a review of recent developments.)

\vspace{-0.2cm}
\section{Simulation}
\label{sec:simu}

\vspace{-0.2cm}
\subsection{Simulation details}
\label{subsec:detail}

In our simulation,
we use the tree level Symanzik gauge action
and the highly improved staggered quark (HISQ) action~\cite{Follana:2006rc}
without the tadpole improvement 
and the mass correction in the Naik term.
It is expected that
the flavor symmetry in the staggered fermion 
and the behavior towards the continuum limit are improved
by HISQ action.
We carried out the simulation
by using the standard Hybrid Monte-Carlo (HMC) algorithm.
We computed the hadron spectrum
as the global survey in the parameter region
and 
we obtained $M_\pi$, $M_\rho$, $f_\pi$	
and $\langle \bar{\psi}\psi \rangle$
as the basic observable.

The simulation for $N_f=4$ is carried out 
at $\beta(=6/g^2)$= 3.6, 3.7 and 3.8
for various quark masses
on $12^3 \times 16$ and $16^3 \times 24$.
We took over 1000 trajectories on the small lattice
and about 600 trajectories on the large lattice
in $N_f=4$ case.
The simulation for $N_f=8$ is carried out 
at $\beta(=6/g^2)$=3.6, 3.7, 3.8, 3.9 and 4.0
for various quark masses
on ($12^3 \times 32$,) $18^3 \times 24$,  $24^3 \times 32$, $30^3 \times 40$ 
and $36^3 \times 48$
for various quark masses.
We took about 800 trajectories on each size.

\vspace{-0.2cm}
\subsection{Analysis methods}
\label{subsec:analysis}

In this section,
we show preliminary results
of  the chiral perturbation analysis  (ChPT) 
and the finite-size hyperscaling analysis (FSHS)
in 4- and 8-flavor case.

If the system is in the chiral symmetry broken phase ($\chi$SB),
physical quantities, $M_H$,  
are described by the chiral perturbation theory (ChPT);
the polynomial behavior.
In particular, about the pion decay constant 
$f_\pi = F + c_1 m_f + c_2 m_f^2 + \cdots. $
(Here we don't discuss the existence of the chiral log.)
If $ F \neq 0$ in the above Eq., it is regarded as the $\chi$SB.

On the other hand,
if the system is in the conformal window,
$M_H$ are described by the finite-size hyperscaling relation (FSHS)~\cite{DelDebbio:2010ze} ; 
$L M_H = {\cal F}(X)$  where $X=L m_f^{\frac{1}{1+\gamma}}$.
The $\gamma$ in this equation is defined as the anomalous mass-dimension.
We carry out the hyperscaling analysis with our data of $M_H=\{M_\pi, f_\pi, M_\rho\}$  
by the following fit function;
$L M_H =  c_0 + c_1 X $.

In the following,
we analyze $N_f=4$ and 8 by these methods.

\vspace{-0.35cm}
\section{Spectrum}
\label{sec:spectrum}

\vspace{-0.2cm}
\subsection{$N_f=4$}
\label{sec:nf4}

\begin{figure}[!h]
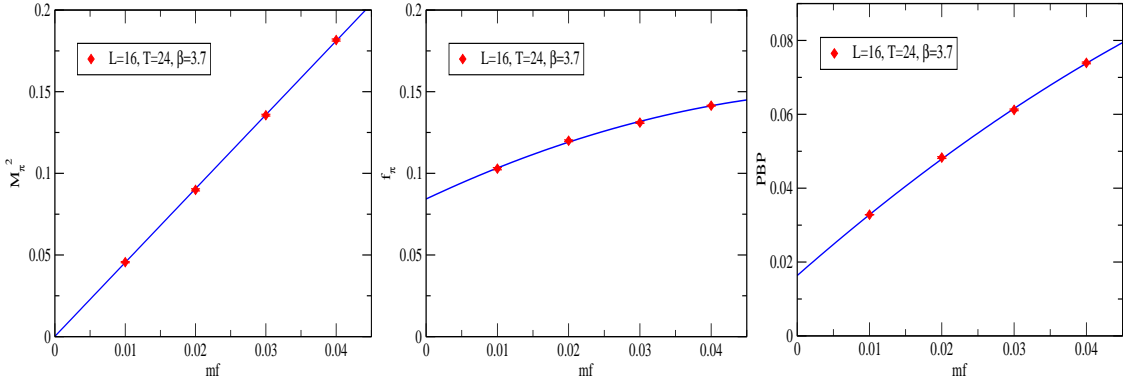
 
\center{
\includegraphics[width=4.9cm,height=5.0cm,trim=0 0 0 0,clip=true]{Figs/Nf04L16T24B3.7mpi2.eps}
\includegraphics[width=4.9cm,height=5.0cm,trim=0 0 0 0,clip=true]{Figs/Nf04L16T24B3.7fpi.eps}
\includegraphics[width=4.9cm,height=5.0cm,trim=0 0 0 0,clip=true]{Figs/Nf04L16T24B3.7PBP.eps}
}
\caption{
In $N_f=4$ $SU(3)$ gauge theory on $16^3 \times 24$ at $\beta=3.7$; 
{\bf Left:} $M_\pi^2$ as functions of $m_f$,
{\bf Center:} $f_\pi$ as functions of $m_f$,
{\bf Right:} $\langle \bar{\psi}\psi \rangle$ as functions of $m_f$.
The blue-solid line in the left panel is the linear fit 
and the lines in other panels are the quadratic fit.
}
\label{fig:nf4}
\end{figure}

\begin{figure}[!h]
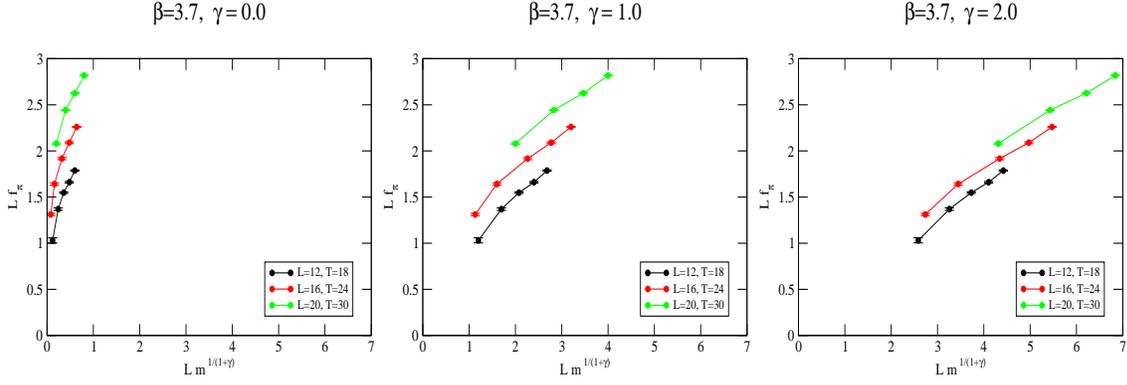
 
\center{
\includegraphics[width=4.9cm,height=5.0cm,trim=0 0 0 0,clip=true]{Figs/fpi_B3.7_g0.0.eps}
\includegraphics[width=4.9cm,height=5.0cm,trim=0 0 0 0,clip=true]{Figs/fpi_B3.7_g1.0.eps}
\includegraphics[width=4.9cm,height=5.0cm,trim=0 0 0 0,clip=true]{Figs/fpi_B3.7_g2.0.eps}
}
\caption{
Finite size hyperscaling test of $f_\pi$ in $N_f=4$ $SU(3)$ gauge theory;
{\bf Left:} at $\gamma=0.0$,
{\bf Center:} at $\gamma=1.0$,
{\bf Right:} at $\gamma=2.0$.
}
\label{fig:nf4fpihs}
\end{figure}

In this subsection,
we analyze $N_f=4$ system
by the ChPT and the finite-size hyperscaling relation.
The result of  $N_f=4$ is shown in Fig.~\ref{fig:nf4},
in which the pion mass squared, the decay constant and the chiral condensate 
are plotted on the panel from the left to the right respectively.
$M_\pi^2$ is proportional to  $m_f$.
$f_\pi$ and $\langle \bar{\psi}\psi \rangle$ 
have the non zero value 
in the chiral limit.
Thus,  the $N_f=4$ has the property of  the $\chi$SB phase
and this is regarded as the signal of the chiral broken phase
in the dynamical case of lattice QCD.
Also, if the FSHS test is applied to $N_f=4$ 
which is $\chi$SB phase (the ordinary QCD),
what happens? 
The result of  this attempt for $f_\pi$ is shown in Fig.~\ref{fig:nf4fpihs}.
From these Figs., 
there is no data alignment in the region $0 \leq \gamma \leq 2$.
This is the property of QCD
when the finite size hyperscaling is applied to.
These properties (ChPT and FSHS ) in $N_f=4$
may hint whether $N_f=8$ is $\chi$SB/walking or conformal.

\vspace{-0.2cm}
\subsection{$N_f=8$}
\label{subsec:nf8}

\begin{figure}[!h]
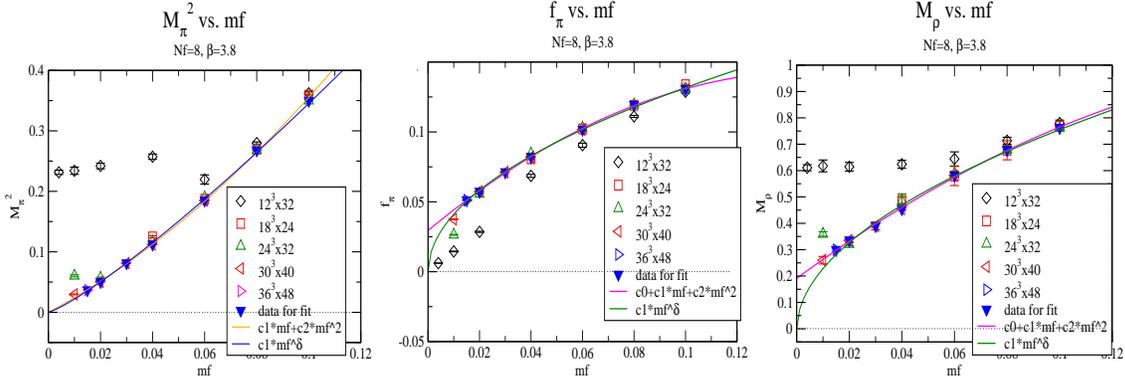
 
\center{
\includegraphics[width=4.9cm,height=5.0cm,trim=0 0 0 0,clip=true]{Figs/fig-fit-mpi2-b3.8.eps}
\includegraphics[width=4.9cm,height=5.0cm,trim=0 0 0 0,clip=true]{Figs/fig-fit-fpi-b3.8.eps}
\includegraphics[width=4.9cm,height=5.0cm,trim=0 0 0 0,clip=true]{Figs//fig-fit-mrho-b3.8.eps}
}
\caption{
ChPT (the quadratic fit) in $f_\pi$ in $N_f=8$ $SU(3)$ gauge theory;
{\bf Left:}  $M_\pi^2$ as a function of $m_f$,
{\bf Center:}  $f_\pi$,
{\bf Right:} $M_\rho$.
The solid lines are the quadratic fit and the power fit.}
\label{fig:nf8chpt}
\end{figure}

In this subsection,
we analyze $N_f=8$ system
by the ChPT and the FSHS test.
The panels in Fig.~\ref{fig:nf8chpt} are $M_\pi$, $f_\pi$ and $M_\rho$ 
at $\beta=3.8$ in particular
as a function of the quark mass $m_f$,
and the polynomial fit (quadratic fit) and the power fit
are plotted.

In $M_\pi$ and $M_\rho$,
the plateau appears  in $m_f \lesssim 0.06$
on small lattice ($12^3 \times 32$)
and  in $m_f \lesssim 0.02$ on large lattice ($24^3 \times 32$) at $\beta=3.8$.
In the corresponding region of the  plateau,
$f_\pi$ behaves like the linear toward the zero.
Since these might be the effect of the finite size effect or might be in a different vacuum,
these data are not included in the following analyses.

In Fig.~\ref{fig:nf8chpt}, 
to take the infinite volume limit is difficult.
Then we take the data on the largest volume at each $m_f$
for the fitting. 
The fit range is $0.0 \leq m_f \leq 0.1$.
The fit result of $M_\pi^2$ is obtained as follows:
$M_\pi^2 = 2.31(2)  m_f + 12.5(1)  m_f^2, (\chi^2/{\rm dof}=17.9)$
and $M_\pi^2 = 5.43(4)  m_f^{1.197(3)} , (\chi^2/{\rm dof}=34.0)$.
For $f_\pi$, 
$\chi^2(f_\pi)/{\rm dof}=14.7$ in the power fit
and $\chi^2(f_\pi)/{\rm dof}=6.1$ in the quadratic fit,
then $f_\pi \rightarrow F = 0.0295(3)$ in the limit $m_f \rightarrow 0$
as the quadratic fit result.
For $M_\rho$,
$\chi^2(M_\rho)/{\rm dof}=6.5$ in the power fit
and  $\chi^2(M_\rho)/{\rm dof}=1.3$ in the quadratic fit, 
then $M_\rho =0.191(8)$ in the limit $m_f \rightarrow 0$
as the the quadratic fit result.
In all cases,
since the $\chi^2/{\rm dof}$ in the quadratic fit is better than that in the power fit,
the chiral limit by the quadratic fit gives the non-zero value of $f_\pi$ and $M_\rho$.
Thus, it seems that the $N_f=8$ is in the $\chi$SB phase.

Here we discuss the validity of the ChPT fit.
We used the expansion parameter defined as
${\mathcal X} = N_f \left( \frac{M_\pi(m_f)}{4 \pi F} \right)^2$,
where $F$ is the pion decay constant in the chiral limit
and $M_\pi(m_f)$ is the pion mass at $m_f$.
In our simulation, ${\cal X} \simeq 1.2-2.5 = O(1)$ 
at the minimum value of $M_\pi \simeq 0.2$
in our simulation.
Therefore, our result in $N_f=8$  is consistent with ChPT.

Next we consider the chiral condensate by the direct calculation, 
$\langle {\bar \Psi} \Psi \rangle={\rm Tr} [D^{-1}_{HISQ}(x,x)]$,
and Gell-Mann-Oakes-Renner (GMOR) relation, $\Sigma=f_\pi^2 M_\pi^2/ (4 m_f)$.
In the result of the direct calculation,
$\langle {\bar \Psi} \Psi \rangle \sim m_f/2$ and then the chiral limit is very small in the lattice unit.
In GMOR relation,
the solid line is the combination
of the quadratic fit results of $M_\pi^2$ and $f_\pi$.
This chiral limit is $\Sigma \simeq 0.0005$
in good coincidence with the direct measurement.
Thus  the chiral limit of the chiral condensate is very small in the lattice unit.
This means the chiral condensate of the techni-quark, $\langle {\bar Q} Q \rangle$,
is tiny value
compared with the mass deformation $m_f$.
Therefore, in the following,
we attempt to find the tail (the remnant) of the conformal.
\begin{figure}[!h]
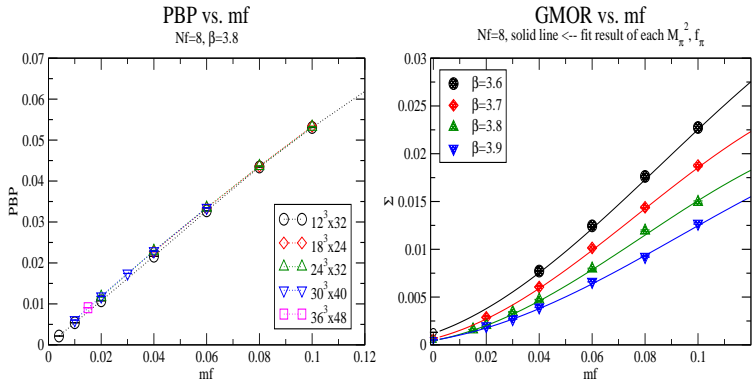
 
\center{
\includegraphics[width=4.9cm,height=5.0cm,trim=0 0 0 0,clip=true]{Figs/fig-pbp-b3.8.eps}
\includegraphics[width=4.9cm,height=5.0cm,trim=0 0 0 0,clip=true]{Figs/fig-gmor.eps}
}
\caption{
Chiral condensate  in $N_f=8$ $SU(3)$ gauge theory;
{\bf Left:}  $\langle {\bar \Psi} \Psi \rangle$ at $\beta=3.8$.
{\bf Right:}  GMOR relation and the line obtained from the quadratic fit
results of $M_\pi^2$ and $f_\pi$,
at various $\beta$s.
}
\label{fig:nf8cond}
\end{figure}

\begin{figure}[!h]
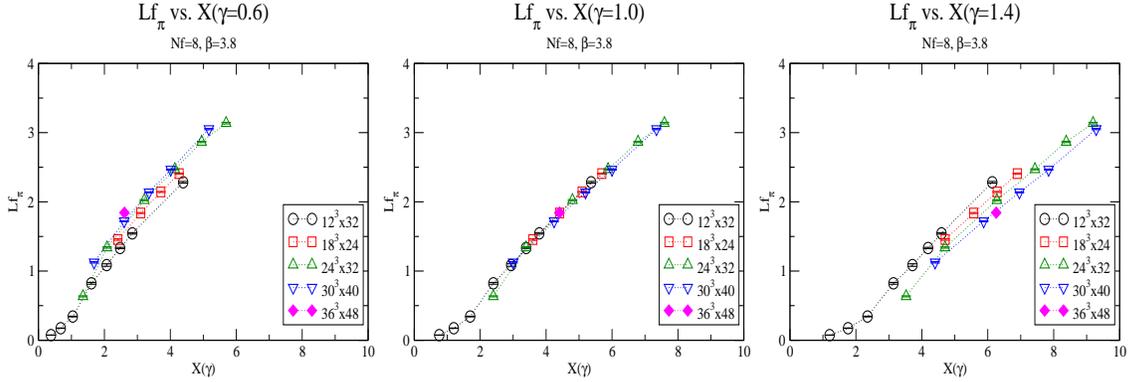
 
\center{
\includegraphics[width=4.9cm,height=5.0cm,trim=0 0 0 0,clip=true]{Figs/fig-fpi-g0.6.eps}
\includegraphics[width=4.9cm,height=5.0cm,trim=0 0 0 0,clip=true]{Figs/fig-fpi-g1.0.eps}
\includegraphics[width=4.9cm,height=5.0cm,trim=0 0 0 0,clip=true]{Figs/fig-fpi-g1.4.eps}
}
\caption{
Finite size hyperscaling test of $f_\pi$ in $N_f=8$ $SU(3)$ gauge theory;
{\bf Left:} at $\gamma=0.6$,
{\bf Center:} at $\gamma=1.0$,
{\bf Right:} at $\gamma=1.4$.
}
\label{fig:nf8fpihs}
\end{figure}

If the system is in the conformal window,
the data are in the good agreement with the FSHS having the universal value of $\gamma$.
Because of the tiny value of the chiral condensate,
we apply the FSHS to $N_f=8$ system in order to catch the tail of the conformal
if there is.
Fig.~\ref{fig:nf8fpihs} is the FSHS test of $f_\pi$ 
for the various $\gamma$.
The data is aligned (collapsing) at around $\gamma=1$.

\begin{figure}[!h]
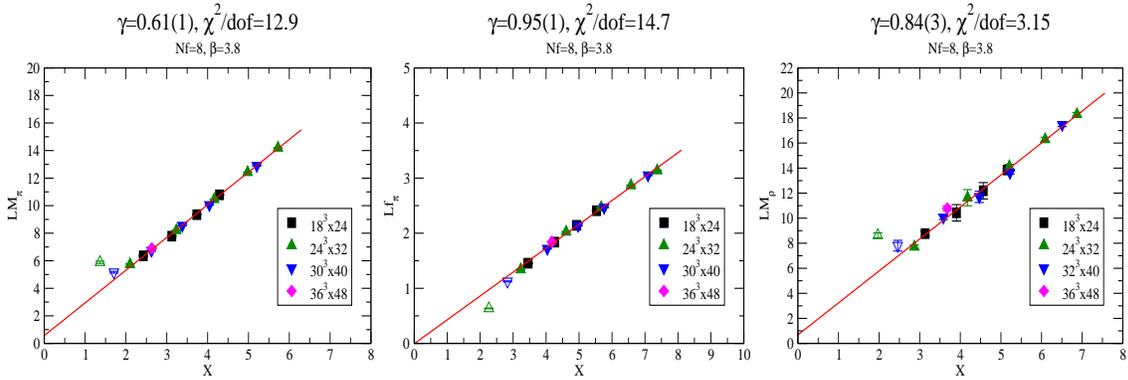
 
\center{
\includegraphics[width=4.9cm,height=5.0cm,trim=0 0 0 0,clip=true]{Figs/fig-hyper-mpi-b3.8.eps}
\includegraphics[width=4.9cm,height=5.0cm,trim=0 0 0 0,clip=true]{Figs/fig-hyper-fpi-b3.8.eps}
\includegraphics[width=4.9cm,height=5.0cm,trim=0 0 0 0,clip=true]{Figs/fig-hyper-mrho-b3.8.eps}
}
\caption{
Finite size hyperscaling fit by the linear ansatz of $f_\pi$ in $N_f=8$ $SU(3)$ gauge theory;
{\bf Left:}  $M_\pi$,
{\bf Center:}  $f_\pi$,
{\bf Right:} $M_\rho$.
The open symbol is not included in the fit data.}
\label{fig:nf8fpihsfit}
\end{figure}

Therefore, 
to quantify this alignment,
we attempt the linear fit as the leading approximation of FSHS.
Panels in Fig.~\ref{fig:nf8fpihsfit} are
the linear fit result of FSHS for 
$M_\pi$, $f_\pi$ and $M_\rho$
from the left to the right.
Although $\chi^2/{\rm dof}$ is not small,
the linearity for each observable is reasonably well
and it seems that the remnant of conformal exists.
The result,  $\gamma(M_\pi) \neq \gamma(M_\rho) \neq \gamma(f_\pi) \sim 1.0$,
indicates the remnant of the conformal.
This situation is very interesting to construct the walking model.

\vspace{-0.2cm}
\section{Discussion and Summary}
\label{sec:summary}

We have made simulations of lattice QCD
with 4 and 8 flavors 
by using the HISQ action.
We obtained the following {\it preliminary} result;
The $N_f=4$ QCD is in good agreement with the chiral broken phase
and the $f_\pi$ data is not aligned in $0 \leq \gamma \leq 2$,
which is the characteristics of QCD applied to the FSHS test.
The $N_f=8$ is consistent with ChPT (${\cal X}$-parameter) 
and is not inconsistent with FSHS
particularly for the data of the relatively large $m_f (> F)$. 
We extracted the $\gamma$-value from the FSHS test;
non-universal $\gamma$ and $\gamma(f_\pi) \sim 1.0$.
We show the table for various $\beta$s as the very preliminary result
(which is revised from that in Ref.~\cite{Aoki:2012ep}).
\begin{table}[!h]
\begin{center}
\begin{tabular}{|c|c|c|c|c|c|}
\hline
         & $\beta=3.6$ & $\beta=3.7$ & $\beta=3.8$ & $\beta=3.9$  & $\beta=4.0$ \\
\hline
 $\gamma$ in $M_\pi$  & 0.64(1)  & 0.63(1)  & 0.61(1)   & 0.56(1)  & 0.56(1) \\
\hline 
 $\gamma$ in $f_\pi$  & 0.98(2)   & 0.99(1)   & 0.95(1)    & 0.92(1) & 0.91(1) \\
\hline
 $\gamma$ in $M_\rho$ & 1.02(2) & 0.91(4) & 0.84(3)  & 0.79(4)  & 0.77(6) \\
\hline 
\end{tabular}
\end{center}
\caption{Preliminary. The statistical error only }
\end{table}
To understand the behavior of $N_f=8$,
we  compare with our simulation of $N_f=12$ 
which is consistent with the conformal~\cite{Aoki:2012eq}
and the Schwinger-Dyson equation analysis on the finite size and mass~\cite{Aoki:2012ve}.
According to SD-eq. analysis,
it is $\gamma \simeq 1.0$ for the near conformal in $\chi$SB phase.
This might indicate that the system with $\gamma \sim1.0$ is near conformal (walking).
Therefore, 
$N_f=8$  would be a good candidate of the walking theory.

We should mention, however,  
that there are several possible systematic uncertainties not considered 
in this report;
As pointed out in Ref.~\cite{Aoki:2012ve},
there exists the mass correction in the hyperscaling relation
for the heavy quark region.
Then the linear ansatz adopted in Fig.~\ref{fig:nf8fpihsfit}  may not be sufficient
to fit our data.
To improve the situation for better understanding,
we will accumulate more data
for various fermion masses and $\beta$s
on larger lattices,
and  carry out detailed analysis using those data.

\vspace{-0.2cm}
\acknowledgments
Numerical simulation has been carried out on 
the supercomputer system $\varphi$
 at KMI in Nagoya university.
This work is supported
by the JSPS Grant-in-Aid for Scientific Research (S) No.22224003, 
 (C) No.23540300 (K.Y.) and (C) No.21540289 (Y.A.),
and also by Grants-in-Aid of the Japanese Ministry for Scientific Research 
on Innovative Areas No. 23105708 (T.Y.). 

\vspace{-0.2cm}


\begin{thebibliography}{99}

\bibitem{Yamawaki:1985zg}
  K.~Yamawaki, M.~Bando and K.~-i.~Matumoto,
  Phys.\ Rev.\ Lett.\  {\bf 56} (1986) 1335.
  

\bibitem{Akiba:1985rr}
Subsequent similar works without notion of anomalous dimension and scale symmetry were done:
  T.~Akiba and T.~Yanagida,
  Phys.\ Lett.\ B {\bf 169} (1986) 432; 
  T.~W.~Appelquist, D.~Karabali and L.~C.~R.~Wijewardhana,
  Phys.\ Rev.\ Lett.\  {\bf 57} (1986) 957.
See also the earlier work on pure numerical analysis:
  B.~Holdom,
  Phys.\ Lett.\ B {\bf 150} (1985) 301.
  
 
 \bibitem{beta} 
  W.~E.~Caswell,
  Phys.\ Rev.\ Lett.\  {\bf 33} (1974) 244; 
  T.~Banks and A.~Zaks,
  Nucl.\ Phys.\ B {\bf 196} (1982) 189.

  
\bibitem{Maskawa:1974vs}
  T.~Maskawa and H.~Nakajima,
  Prog.\ Theor.\ Phys.\  {\bf 52} (1974) 1326; 
  R.~Fukuda and T.~Kugo,
  Nucl.\ Phys.\ B {\bf 117} (1976) 250;
  

\bibitem{Farhi:1980xs}
See for a review: 
  E.~Farhi and L.~Susskind,
  Phys.\ Rept.\  {\bf 74} (1981) 277.
  
    

\bibitem{Iwasaki:1991mr}
  Y.~Iwasaki, K.~Kanaya, S.~Sakai and T.~Yoshie,
  Phys.\ Rev.\ Lett.\  {\bf 69} (1992) 21;
  Y.~Iwasaki, K.~Kanaya, S.~Kaya, S.~Sakai and T.~Yoshie,
  Phys.\ Rev.\  D {\bf 69} (2004) 014507
  [arXiv:hep-lat/0309159].

\bibitem{Appelquist:2007hu}
  T.~Appelquist, G.~T.~Fleming and E.~T.~Neil,
  Phys.\ Rev.\ Lett.\  {\bf 100} (2008) 171607
  [Erratum-ibid.\  {\bf 102} (2009) 149902]
  [arXiv:0712.0609 [hep-ph]];
  T.~Appelquist, G.~T.~Fleming and E.~T.~Neil,
  Phys.\ Rev.\  D {\bf 79} (2009) 076010
  [arXiv:0901.3766 [hep-ph]].


\bibitem{plenary}
J.~Giedt,
PoS LATTICE {\bf 2012} (2012) 006;
  E.~T.~Neil,
  PoS LATTICE {\bf 2011} (2011) 009
  [arXiv:1205.4706 [hep-lat]].
References therein.


\bibitem{Follana:2006rc}
  E.~Follana {\it et al.}  [HPQCD Collaboration and UKQCD Collaboration],
  Phys.\ Rev.\  D {\bf 75} (2007) 054502
  [arXiv:hep-lat/0610092];
  A.~Bazavov {\it et al.}  [MILC collaboration],
  Phys.\ Rev.\  D {\bf 82} (2010) 074501
  [arXiv:1004.0342 [hep-lat]].


\bibitem{DelDebbio:2010ze}
  L.~Del Debbio and R.~Zwicky,
  Phys.\ Rev.\  D {\bf 82} (2010) 014502
  [arXiv:1005.2371 [hep-ph]].

  
\bibitem{Aoki:2012ep}
  Y.~Aoki, T.~Aoyama, M.~Kurachi, T.~Maskawa, K.~-i.~Nagai, H.~Ohki, A.~Shibata and K.~Yamawaki and T.~Yamazaki  [LatKMI Collaboration],
  PoS LATTICE {\bf 2011} (2011) 080
  [arXiv:1202.4712 [hep-lat]].
  
\bibitem{Aoki:2012eq}
  Y.~Aoki, T.~Aoyama, M.~Kurachi, T.~Maskawa, K.~-i.~Nagai, H.~Ohki, A.~Shibata and K.~Yamawaki  and T.~Yamazaki  [LatKMI Collaboration],
  Phys.\ Rev.\ D {\bf 86} (2012) 054506
  [arXiv:1207.3060 [hep-lat]]; 
PoS LATTICE {\bf 2012} (2012) 029 
[arXiv:1211.6651 [hep-lat]].
  
  
\bibitem{Aoki:2012ve}
  Y.~Aoki, T.~Aoyama, M.~Kurachi, T.~Maskawa, K.~-i.~Nagai, H.~Ohki, A.~Shibata and K.~Yamawaki and T.~Yamazaki  [LatKMI Collaboration],
  Phys.\ Rev.\ D {\bf 85} (2012) 074502
  [arXiv:1201.4157 [hep-lat]]; 
 PoS LATTICE {\bf 2012} (2012) 059. 
 
  

\end{thebibliography}
\end{document}